\documentclass[prd,showpacs,amsmath,amssymb,nobibnotes,nofootinbib,11pt]{revtex4}
\usepackage{latexsym}
\usepackage{bm}

\begin{document}

\title{A note on cosmological Levi-Civita spacetimes in higher dimensions}

\author{{\"O}zg{\"u}r Sar{\i}o\u{g}lu}
\email{sarioglu@metu.edu.tr}
\affiliation{Department of Physics, Faculty of Arts and  Sciences,\\
             Middle East Technical University, 06531, Ankara, Turkey}

\author{Bayram Tekin}
\email{btekin@metu.edu.tr}
\affiliation{Department of Physics, Faculty of Arts and  Sciences,\\
             Middle East Technical University, 06531, Ankara, Turkey}

\date{\today}

\begin{abstract}
We find a class of solutions to cosmological Einstein equations that generalizes
the four dimensional cylindrically symmetric spacetimes to higher dimensions. The
AdS soliton is a special member of this class with a unique singularity structure.
\end{abstract}


\maketitle

In this note, we generalize the four dimensional cylindrically symmetric
static spacetimes that are solutions to the Einstein equations, with a negative 
cosmological constant $\Lambda<0$, to higher dimensions. The flat space ($\Lambda=0$) 
analogs of these solutions are known as the Levi-Civita spacetimes \cite{mac}.

The original four dimensional solutions with a non-vanishing $\Lambda$ were found 
in \cite{linet} and \cite{tian}. Nonsingular sheet sources for these spacetimes 
were recently constructed in \cite{zofbic}. There are subtle issues in the 
interpretation of the parameters that appear in the four dimensional solution in 
terms of the physical properties, such as the mass, of material sources. According 
to \cite{zofbic}, ``for the $\Lambda<0$ case the asymptotic forms of
the metrics due to material cylinders are more closely related to the
asymptotics of bounded sources than for the $\Lambda=0$ case.''

Here we provide a new class of solutions to the cosmological Einstein equations (with
a redefined cosmological constant)
\[ R_{\mu\nu} = - \frac{n}{\ell^{2}} g_{\mu\nu} \]
in $D=n+1 \geq 4$ dimensions. It reads
\begin{equation}
ds^{2} = \frac{r^2}{\ell^2} \left[ - \left( 1 - \frac{r_0^{n}}{r^{n}} \right)^{p_{0}} dt^2 
+ \sum_{k=1}^{n-1} \left( 1 - \frac{r_0^{n}}{r^{n}} \right)^{p_{k}} (dx^k)^{2} \right] + 
\left( 1 - \frac{r_0^{n}}{r^{n}} \right)^{-1} \, \frac{\ell^2}{r^2} \, dr^2 \,,
\label{sol}
\end{equation}
where $r_{0}$ is a free parameter, and the constants $p_{k}$, with $0 \leq k \leq n-1$,
satisfy two Kasner-type conditions:
\begin{equation}
\sum_{k=0}^{n-1} p_{k} = \sum_{k=0}^{n-1} (p_{k})^{2} = 1 \, . \label{cons}
\end{equation}
This is a higher dimensional generalization of the metrics in \cite{linet, tian}. 
Specifically, the metric studied by \cite{zofbic} follows from (\ref{sol}) after 
solving the constraints (\ref{cons}) for $n=3$ \footnote{Their parameter $\sigma$ 
can be thought of as the remaining unconstrained $p$.}.

It immediately follows that as $r \to \infty$, (\ref{sol}) asymptotically approaches 
the usual maximally symmetric AdS spacetime in horospheric coordinates. Another important 
observation is that the AdS soliton of \cite{horo} is just a special member in this 
class: It is obtained simply by taking one of the $p_{k}=1$, where $1 \leq k \leq n-1$, and 
setting all the others, including $p_{0}$, to zero. In lower dimensions, when $n=1$ and $n=2$, 
(\ref{sol}) becomes the usual two and three dimensional AdS spacetimes, respectively. Here we 
should also note that special forms of (\ref{sol}) with specific choices for the constants 
$p_{k}$ have already appeared in the literature for $D=5$ and $D=7$ \cite{myers,russo,park}. 

The singularity structure of the solution (\ref{sol}) is apparent from its Kretschmann scalar,
which we give here only for the $n=3$ and $n=4$ cases \footnote{Its calculation gets rather 
complicated beyond these dimensions, but is not different in general features.}. For $n=3$, 
it reads
\[ R_{\mu\nu\lambda\sigma} R^{\mu\nu\lambda\sigma} = 
\frac{24}{\ell^{4}} + \frac{12}{\ell^{4}} \frac{r_{0}^{6}}{r^{6}} - \frac{81}{\ell^{4}} 
\frac{r_{0}^{9}}{r^{9}} \, \frac{(1+h(r))}{h(r)^{2}} \, \prod_{k=0}^{2} p_{k} \, , \]
where $h(r) \equiv 1-r_{0}^{3}/r^{3}$. As for the $n=4$ case, it will be convenient to first
define $f(r) \equiv 1-r_{0}^{4}/r^{4}$ and 
$\Delta \equiv p_{0} p_{1} p_{2} + p_{0} p_{1} p_{3} + p_{0} p_{2} p_{3} + p_{1} p_{2} p_{3}$. Then
\[ R_{\mu\nu\lambda\sigma} R^{\mu\nu\lambda\sigma} = 
\frac{40}{\ell^{4}} + \frac{72}{\ell^{4}} \frac{r_{0}^{8}}{r^{8}} - \frac{64}{\ell^{4}} 
\frac{r_{0}^{12}}{r^{12}} \, \frac{1}{f(r)^{2}} \, \Big( \Delta (4+5f(r)) 
+ 2 \, \frac{r_{0}^{4}}{r^{4}} \, \prod_{k=0}^{3} p_{k} \Big) \, . \]
For the generic choice of the constants $p_{k}$, one clearly finds naked singularities at 
$r=r_{0}$ and $r=0$. Remarkably, the AdS soliton is the unique solution with no naked
singularities provided that $r \geq r_{0}$ and the conical singularity at $r_{0}$ is
avoided by a proper compactification of the corresponding coordinate \cite{horo}.

It may be of interest to write (\ref{sol}) in different coordinate systems. A
somewhat obvious possibility is to consider the coordinate transformation
\( r = r_{0} \cosh^{2/n}(n \rho/(2\ell)) \) which takes (\ref{sol}) to 
\begin{equation}
ds^{2} = d \rho^{2} + \frac{r_{0}^{2}}{\ell^{2}} \, \cosh^{4/n}\Big( \frac{n \rho}{2\ell} \Big) \, 
\Big(- \tanh^{2 p_{0}}\Big( \frac{n \rho}{2\ell} \Big) dt^{2} + 
\sum_{k=1}^{n-1} \, \tanh^{2 p_{k}}\Big( \frac{n \rho}{2\ell} \Big) \, (dx^k)^{2} \Big)\,,
\label{newads}
\end{equation}
where again the constants $p_{k}$ are subject to (\ref{cons}), of course.

The interpretation of the constants $p_{k}$ and $r_{0}$ in terms of physical parameters
would be of interest. As a step in this direction, we calculate the mass of the solutions 
(\ref{sol}). For this purpose, we use the procedure described in \cite{dt1, dt2} that
requires a choice of a background and a perturbation about it. The correct background 
in this case is the usual AdS metric obtained from (\ref{sol}) by simply setting $r_{0}$ 
to zero. The background Killing vector that leads to mass/energy 
is $\bar{\xi}^{\mu} = - (\partial/\partial t)^{\mu}$, which in our case yields
\begin{equation} 
M = \frac{V_{n-1}}{4 \, G_{D} \, \Omega_{n-2}} \, \frac{r_{0}^{n}}{\ell^{n+1}} \, (n p_{0} - 1) \,, \label{yuk}
\end{equation}
where $V_{n-1}$ is the volume of the transverse dimensions. The mass of the AdS soliton, 
for which there is only one non-zero $p_{k}$ where $k \neq 0$, has already been 
considered in \cite{horo,cst}. It was conjectured in \cite{horo} that the AdS soliton, 
with its negative mass, has the lowest possible energy, (hence the new `positive' 
mass conjecture) in its asymptotic class. At first sight, the result (\ref{yuk}) 
seems to contradict this since one can have a solution with $p_{0}<0$ leading to
\( M < M_{soliton} \). However, one then has naked singularities in which case any kind of 
`positive' mass theorem fails. Observing that the AdS soliton is the unique solution 
with no naked singularities, our result lends further support to the new `positive'
mass conjecture of \cite{horo} and the ``uniqueness'' result of \cite{gsw}.
Amusingly, if one chooses $p_{0}=1/n$, one ends up with massless non-flat $(n-1)$-branes.

To conclude, we have found new Einstein spaces in higher dimensions, generalizing 
the four dimensional cosmological cylindrically symmetric Levi-Civita spacetimes. 
We have calculated the mass of these metrics and identified the AdS soliton as a rather
unique member.

\section{\label{ackno} Acknowledgments}
This work is partially supported by the Scientific and Technological Research
Council of Turkey (T{\"U}B\.{I}TAK). B.T. is also partially supported by the 
T{\"U}B\.{I}TAK Kariyer Grant 104T177.

\end{document}